\documentclass[11pt,epsf,letterpaper]{article}%
\usepackage{subfig}
\usepackage{color}
\usepackage{amsmath}
\usepackage{amsfonts}
\usepackage{verbatim}
\usepackage{amssymb}
\usepackage{graphicx}
\usepackage{epstopdf}
\usepackage{mathrsfs}
\usepackage{epsfig}%
\providecommand{\U}[1]{\protect\rule{.1in}{.1in}}

\begin{document}

\title{On the thermodynamics of hairy black holes}
\author{$^{(1)}$Andr\'{e}s Anabal\'{o}n, $^{(2)}$Dumitru Astefanesei and $^{(3)}$David Choque\\
\\\textit{$^{(1)}$Departamento de Ciencias, Facultad de Artes Liberales y} \\\textit{Facultad de Ingenier\'{\i}a y Ciencias, Universidad Adolfo
Ib\'{a}\~{n}ez,}\\\textit{Vi\~{n}a del Mar, Chile.} \\
\\\textit{$^{(2)}$ Instituto de F\'\i sica, Pontificia Universidad Cat\'olica de
Valpara\'\i so,} \\\textit{Casilla 4059, Valpara\'{\i}so, Chile.} \\
\\\textit{$^{(3)}$Universidad T\'{e}cnica Federico Santa Mar\'{\i}a,} \\\textit{Av. Espa\~{n}a 1680, Valparaiso, Chile.}}
\maketitle

\begin{abstract}
We investigate the thermodynamics of a general class of exact 4-dimensional
asymptotically Anti-de Sitter hairy black hole solutions and show that, for a
fixed temperature, there are small and large hairy black holes similar to the
Schwarzschild-AdS black hole. The large black holes have positive specific
heat and so they can be in equilibrium with a thermal bath of radiation at the
Hawking temperature. The relevant thermodynamic quantities are computed by
using the Hamiltonian formalism and counterterm method. We explicitly show
that there are first order phase transitions similar to the Hawking-Page phase
transition.\newline

\end{abstract}

\newpage


\section{Introduction}

Asymptotically Anti-de Sitter (AdS) black holes play an important role in
understanding the dynamics and thermodynamics of holographic dual field
theories via AdS/CFT duality \cite{Maldacena:1997re}. In particular, these
black holes are dual to thermal states of the `boundary' field theory. First
order phase transitions in the bulk can be related to
confinement/deconfinement-like phase transitions in the dual field theory
\cite{Witten:1998zw}. Since scalar fields appear as moduli in string theory,
it is important to understand the generic thermodynamic properties of hairy
black holes.

Motivated by these considerations, in this paper we study in detail the
thermodynamics of a general class of exact $4$-dimensional neutral hairy black
holes \cite{Anabalon:2012ta, Anabalon:2013sra, Anabalon:2013eaa}
(generalizations to other dimensions or for different horizon topologies can
be found in \cite{Feng:2013tza, Anabalon:2013qua, Lu:2014maa, Acena:2012mr,
Lu:2013ura, Acena:2013jya, Xu:2014uka, Wen:2015xea}) . The scalar potential is
characterized by two parameters and the black hole solution has one
integration constant that is related to its mass. For some particular values
of the parameters in the potential, the solutions can be embedded in
supergravity \cite{Anabalon:2013eaa, Feng:2013tza}. The scalar field potential
contains as special cases all the uncharged exact static solutions so far
discussed in the literature \cite{Martinez:2004nb, Kolyvaris:2009pc,
Gonzalez:2013aca} (for details, see \cite{Anabalon:2012dw}). These static
configurations have been extended to dynamical black hole solutions
\cite{Zhang:2014dfa, Zhang:2014sta}.

There are some subtleties in defining the mass of a hairy black hole
\cite{Barnich:2002pi, Hertog:2004dr, Henneaux:2006hk, Amsel:2006uf,
Anabalon:2014fla}. In \cite{Anabalon:2014fla}, a concrete method of computing
the mass of an asymptotically AdS hairy black hole was proposed. This method
is very useful from a practical point of view because it is using just the
expansions of the metric functions at the boundary. More importantly, it can
be used for hairy black holes that preserve or not the conformal symmetry (the
AdS isometries) of the boundary. We are going to use this method, which is
based on the Hamiltonian formalism \cite{Regge:1974zd}, to compute the mass of
the black hole solutions.

However, based on the physics of AdS/CFT duality, a different method was
developed, the so called `holographic renormalization'
\cite{Henningson:1998gx} (see, also, \cite{Balasubramanian:1999re,
Skenderis:2000in, Skenderis:2002wp, Papadimitriou:2004ap, de Haro:2000xn}%
\footnote{A similar method for asymptotically flat spacetimes was developed in
\cite{Mann:2005yr, Astefanesei:2005ad} and some concrete applications were
presented in \cite{Astefanesei:2009wi, Astefanesei:2010bm}.}) --- for boundary
mixed conditions for the scalar field, this method was further developed in
\cite{Papadimitriou:2007sj, noi}. The main idea behind this method is that,
due to the holography, the infrared (IR) divergences that appear in the
gravity side are equivalent with the ultraviolet divergencies of the dual
field theory. Then, to cure these divergencies, one needs to add counterterms
that are local and depend on the intrinsic geometry of the boundary. In this
way, one can use the quasilocal formalism of Brown and York
\cite{Brown:1992br} supplemented with these counterterms to compute the
regularized Euclidean action and the `boundary' stress tensor. The energy is
the charge associated with the Killing vector $\partial_{t}$ and it can be
obtained from the boundary stress tensor.

Armed with these results, one can investigate the thermodynamics and phase
diagram of the hairy black hole solutions. In particular, we show that there
are first order phase transitions that lead to a discontinuity in the entropy.
Similar results were obtained for a general class of black holes solutions in
a theory with a conformal invariant scalar field Lagrangian
\cite{Giribet:2014fla}.

The rest of the paper is organized as follows: In section $2$, we review the
exact hairy black hole solutions and briefly present some of their properties.
Section $3$ is dedicated to the computations of the thermodynamic quantities.
To gain some intuition, we present a detailed computation of Schwarzschild-AdS
(SAdS) black hole in the coordinates the hairy solution was written. Then, to
regularize the Euclidean action (and the boundary stress tensor), we propose a
counterterm that depends on the scalar field and it is intrinsic to the
boundary. The mass is computed with the counterterm formalism and, also, by
the method of \cite{Anabalon:2014fla}. In section $4$, we use the results in
the previous section to investigate the existence of the phase transitions.
Then, section $5$ concludes with a summary of results.

\section{Black hole solution}

We are interested in asymptotically AdS hairy black hole solutions with a
spherical horizon \cite{Anabalon:2013sra, Anabalon:2013eaa}. The action is
\begin{equation}
I[g_{\mu\nu},\phi]=\int_{\mathcal{M}}{d^{4}x\sqrt{-g}\biggl{[}\frac{R}%
{2\kappa}-\frac{(\partial\phi)^{2}}{2}-V(\phi)\biggr{]}}+\frac{1}{\kappa}%
\int_{\partial\mathcal{M}}{d^{3}xK\sqrt{-h}} \label{action}%
\end{equation}
where $V(\phi)$ is the scalar potential, $\kappa=8\pi G_{N}$, and
the last term is the Gibbons-Hawking boundary term. Here, $h_{ab}$
is the boundary metric and $K$ is the trace of the extrinsic
curvature. The metric ansatz is
\begin{equation}
ds^{2}=\Omega(x)\left[  -f(x)dt^{2}+\frac{\eta^{2}dx^{2}}{f(x)}+d\theta
^{2}+\sin^{2}\theta d\phi^{2}\right]  \label{Ansatz}%
\end{equation}
We consider the following scalar potential, which for some particular values
of the parameters it becomes the one of a truncation of $\omega$-deformed
gauged $N=8$ supergravity \cite{Anabalon:2013eaa, Guarino:2013gsa,
Tarrio:2013qga}:
\begin{align}
V(\phi)  &  =\frac{\Lambda(\nu^{2}-4)}{6\kappa\nu^{2}}\biggl{[}\frac{\nu
-1}{\nu+2}e^{-\phi l_{\nu}(\nu+1)}+\frac{\nu+1}{\nu-2}e^{\phi l_{\nu}(\nu
-1)}+4\frac{\nu^{2}-1}{\nu^{2}-4}e^{-\phi l_{\nu}}\biggr{]}\\
&  +\frac{\alpha}{\kappa\nu^{2}}\biggl{[}\frac{\nu-1}{\nu+2}\sinh{\phi l_{\nu
}(\nu+1)}-\frac{\nu+1}{\nu-2}\sinh{\phi l_{\nu}(\nu-1)}+4\frac{\nu^{2}-1}%
{\nu^{2}-4}\sinh{\phi l_{\nu}}\biggr{]}\nonumber
\end{align}
The equations of motion can be integrated for the conformal factor
\cite{Anabalon:2013sra, Anabalon:2013qua, Acena:2012mr,Acena:2013jya}:
\begin{equation}
\Omega(x)=\frac{\nu^{2}x^{\nu-1}}{\eta^{2}(x^{\nu}-1)^{2}} \label{omega}%
\end{equation}
where $\alpha$ and $\nu$ are two parameters that characterize the hairy
solution. With this choice of the conformal factor, it is straightforward to
obtain the expressions for the scalar field
\begin{equation}
\phi(x)=l_{\nu}^{-1}\ln{x}%
\end{equation}
and metric function
\begin{equation}
f(x)=\frac{1}{l^{2}}+\alpha\biggl{[}\frac{1}{\nu^{2}-4}-\frac{x^{2}}{\nu^{2}%
}\biggl{(}1+\frac{x^{-\nu}}{\nu-2}-\frac{x^{\nu}}{\nu+2}%
\biggr{)}\biggr{]}+\frac{x}{\Omega(x)} \label{f}%
\end{equation}
where $\eta$ is the only integration constant and $l_{\nu}^{-1}=\sqrt{(\nu
^{2}-1)/2\kappa}$.

The potential and the solution are invariant under the transformation
$\nu\rightarrow-\nu$. For $x=1$, which corresponds to the boundary, we can
show that the theory has a standard AdS vacuum $V(\phi=0)=\frac{\Lambda
}{\kappa}$. In the limit $\nu=1$, one gets $l_{\nu}\rightarrow\infty$ and
$\phi\rightarrow0$ so that the SAdS black hole is smoothly obtained.

The mass of the scalar field can be easily computed from the expansion of the
potential and we obtain $m^{2}=-2/l^{2}$, which is the `conformal' mass. It is
also important to point out that there are two distinct branches, one that
corresponds to $x\in[0,1]$ and the other one to $x\in[1,\infty]$ --- the
boundary is at $x=1$ and the curvature singularities are at $x=0$ for the
first branch and $x\rightarrow\infty$ for the second one (these are the
locations where the scalar field is also blowing up).

\section{Thermodynamics}

In this section we use the quasilocal formalism supplemented with counterterms
to compute the Euclidean action and hairy black hole's energy. For
completeness, we also compute the mass with the method of
\cite{Anabalon:2014fla} that is based on the Hamiltonian formalism
\cite{Regge:1974zd}.

As an warm-up example, let us start with SAdS black hole in the coordinates
(\ref{Ansatz}) that can be obtained when the hair parameter is $\nu=1$. The
metric (\ref{Ansatz}) becomes in this case
\begin{equation}
\Omega(x)=\frac{1}{\eta^{2}(x-1)^{2}}\,\,,\qquad f(x)=\frac{1}{l^{2}}+\frac
{1}{3}\alpha(x-1)^{3}+\eta^{2}x(x-1)^{2} \label{schw}%
\end{equation}
To obtain the SAdS black hole in the canonical form, one has to change the
coordinates as
\begin{equation}
x=1\pm\frac{1}{\eta r} \label{change}%
\end{equation}
The reason is that, as we have already discussed, there are two branches that
correspond to $x\in\lbrack0,1]$ and $x\in\lbrack1,\infty]$. Since there are
some subtleties for computing the action for $x\in\lbrack1,\infty]$ branch
(for example, the extrinsic curvature is changing the sign due to a change of
the normal to the foliation $x=$ constant), in what follows we explicitly work
with the branch $x\in\lbrack0,1]$. In this case, using the change of
coordinates (\ref{change}) we obtain the SAdS black hole in canonical
coordinates:
\begin{equation}
\Omega(x)f(x)=F(r)=1-\frac{\mu}{r}+\frac{r^{2}}{l^{2}}\qquad,\qquad\mu
=\frac{\alpha+3\eta^{2}}{3\eta^{3}}%
\end{equation}
It is well known that the action has divergences even at the tree level due to
the integration on an infinite volume. To regularize the action, we use the
counterterms \cite{Balasubramanian:1999re}:
\begin{equation}
I[g_{\mu\nu}]=I_{bulk}+I_{GH}-\frac{1}{\kappa}\int_{\partial\mathcal{M}}%
{d^{3}x\sqrt{-h}\biggl{(}\frac{2}{l}+\frac{\mathcal{R}l}{2}\biggr{)}}%
\end{equation}
where $\mathcal{R}$ is the Ricci scalar of the boundary metric $h_{ab}$.

Let us first compute the bulk action --- in this case, since the scalar field
vanishes the potential becomes the cosmological constant: $V=\frac{\Lambda
}{\kappa}=-\frac{3}{l^{2}\kappa}$. We use the trace of the Einstein tensor and
following combinations of the equations of motion
\begin{align}
E_{t}^{t}-E_{\phi}^{\phi}=0\Rightarrow0  &  =f^{^{\prime\prime}}+\frac
{\Omega^{^{\prime}}f^{^{\prime}}}{\Omega}+2\eta^{2}\\
E_{t}^{t}+E_{\phi}^{\phi}=0\Rightarrow2\kappa V(\phi)  &  =-\frac
{(f\Omega^{^{\prime\prime}}+f^{^{\prime}}\Omega^{^{\prime}})}{\Omega^{2}%
\eta^{2}}+\frac{2}{\Omega}\nonumber
\end{align}
to obtain
\begin{equation}
I_{bulk}^{E}=\frac{4\pi\beta}{\eta^{3}\kappa l^{2}}\biggl{[}-\frac{1}%
{(x_{b}-1)^{3}}+\frac{1}{(x_{h}-1)^{3}}\biggr{]}=\frac{4\pi\beta}{\kappa
l^{2}}(r_{b}^{3}-r_{h}^{3})
\end{equation}
Here, $x_{b}$ and $x_{h}$ are the boundary and horizon locations, and $\beta$
is the periodicity of the Euclidean time that is related to the temperature by
$\beta=T^{-1}$.

The Gibbons-Hawking surface term can be computed if we choose a foliation $x=$
constant with the induced metric $ds^{2}=h_{ab}dx^{a}dx^{b}=\Omega(x)\left[
-f(x)dt^{2}+d\theta^{2}+\sin^{2}\theta d\phi^{2}\right]  $. The normal to the
surface $x=$ constant and extrinsic curvature are
\begin{equation}
n_{a}=\frac{\delta_{a}^{x}}{\sqrt{g^{xx}}}\,\,,\qquad K_{ab}=\frac
{\sqrt{g^{xx}}}{2}\partial_{x}h_{ab}%
\end{equation}
and the contribution of the Gibbons-Hawking term to the action is
\begin{equation}
I_{GH}^{E}=-\frac{2\pi\beta}{\kappa}\biggl{[}-\frac{6}{l^{2}\eta^{3}(x-1)^{3}%
}-\frac{4}{\eta(x-1)}-\biggl{(}\frac{\alpha+3\eta^{2}}{\eta^{3}}%
\biggr{)}\biggr{]}\biggr{\vert}_{x_{b}}=-\frac{2\pi\beta}{\kappa
}\biggl{(}\frac{6r_{b}^{3}}{l^{2}}+4r_{b}-3\mu\biggr{)}
\end{equation}
The last contribution is given by the gravitational counterterm, which is an
intrinsic surface term that depends only on the geometry of the boundary
\begin{equation}
I_{ct}^{E}=\frac{2\pi\beta}{\kappa}\biggl{[}\frac{4}{l^{2}\eta^{3}%
(x_{b}-1)^{3}}+\frac{4}{\eta(x_{b}-1)}-2\mu\biggr{]}=\frac{2\pi\beta}{\kappa
}\biggl{(}\frac{4r_{b}^{3}}{l^{2}}+4r_{b}-2\mu\biggr{)}
\end{equation}
We can explicitly see that the divergences proportional with $r_{b}%
\rightarrow\infty$ and $(r_{b})^{3}\rightarrow\infty$ cancel out and so the
regularized action is
\begin{equation}
I^{E}=I_{bulk}^{E}+I_{GH}^{E}+I_{ct}^{E}=\frac{4\pi\beta}{\kappa l^{2}%
}\biggl{[}\frac{1}{\eta^{3}(x_{h}-1)^{3}}+\frac{\mu l^{2}}{2}\biggr{]}=\frac
{4\pi\beta}{\kappa l^{2}}\biggl{(}-r_{h}^{3}+\frac{\mu l^{2}}{2}\biggr{)}
\end{equation}
The computations for the general hairy black hole (\ref{Ansatz}),
(\ref{omega}), (\ref{f}) are more involved but similar with the ones above and
we do not present all the details here. In this case, the action should be
supplemented with a counterterm that depends also on the scalar field
\cite{Henningson:1998gx, Skenderis:2002wp, Papadimitriou:2007sj,
noi}.\footnote{A more detailed analysis including concrete counterterms for
(non-)logarithmic branch and a comparison with the Hamiltonian formalism is
going to be presented in \cite{noi}.} We work with a counterterm that is
intrinsic to the boundary geometry (it does not depend on the normal to the
boundary or the normal derivatives of the scalar field)
\cite{Papadimitriou:2007sj, noi}:
\begin{equation}
I_{\phi}^{E}=\int_{\partial\mathcal{M}}{d^{3}x^{E}\sqrt{h^{E}}\biggl{(}\frac
{\phi^{2}}{2l}-\frac{l_{\nu}}{6l}\phi^{3}\biggr{)}}=\frac{4\pi\beta}{\kappa
}\biggl{[}-\frac{\nu^{2}-1}{4l^{2}\eta^{3}(x_{b}-1)}+\frac{\nu^{2}-1}%
{3l^{2}\eta^{3}}\biggr{]} \label{Iphi}%
\end{equation}
The sum of the other terms in the action is
\begin{equation}
I_{bulk}^{E}+I_{surf}^{E}+I_{ct}^{E}=-\frac{1}{T}\biggl{(}\frac{AT}%
{4G}\biggr{)}+\frac{4\pi\beta}{\kappa}\biggl{[}\frac{\nu^{2}-1}{4l^{2}\eta
^{3}(x_{b}-1)}+\frac{12\eta^{2}l^{2}+4\alpha l^{2}-4\nu^{2}+4}{12l^{2}\eta
^{3}}\biggr{]}
\end{equation}
where $A=4\pi\Omega(x_{h})$ is the area of the horizon. It is worth mentioning
that the gravitational counterterm \cite{Balasubramanian:1999re} is not
sufficient to cancel the divergence in the action (there is still a term
proportional to $(x_{b}-1)^{-1}$) but when we add the counterterm (\ref{Iphi})
we obtain a finite action:
\begin{equation}
I^{E}=\beta\biggr{(}-\frac{AT}{4G}+\frac{4\pi}{\kappa}\frac{3\eta^{2}+\alpha
}{3\eta^{3}}\biggr{)}
\end{equation}
In the classical limit, the action is related to the thermodynamic potential
(the free energy $F$ in our case), which is $F=I^{E}/\beta=M-TS$. Using the
well known thermodynamic relations or by comparing the two formulas, one can
extract the mass of the hairy black hole:
\begin{equation}
M=\frac{1}{2G}\left(  \frac{\alpha+3\eta^{2}}{3\eta^{3}}\right)
\end{equation}
Since we have constructed the regularized action, we can use the quasilocal
formalism of Brown and York \cite{Brown:1992br} to construct the boundary
stress tensor, which is the variation of the action with respect to the
induced metric:
\begin{equation}
\tau_{ab}=-\frac{1}{\kappa}\biggl{(}K_{ab}-h_{ab}K+\frac{2}{l}h_{ab}%
-lE_{ab}\biggr{)}-\frac{h_{ab}}{l}\biggl{(}\frac{\phi^{2}}{2}-\frac{l_{\nu}%
}{6l}\phi^{3}\biggr{)}
\end{equation}
The boundary metric can be locally written in ADM-like form:
\begin{equation}
h_{ab}dx^{a}dx^{b}=-N^{2}dt^{2}+\sigma_{ij}(dy^{i}+N^{i}dt)(dy^{j}+N^{j}dt)
\end{equation}
where $N$ and $N^{i}$ are the lapse function and the shift vector respectively
and ${y^{i}}$ are the intrinsic coordinates on a (closed) hypersurface
$\Sigma$. The boundary geometry has an isometry generated by the Killing
vector $\epsilon^{a}=(\partial_{t})^{a}$ for which the conserved charge is the
mass:
\begin{equation}
M=Q_{\frac{\partial}{\partial t}}=\oint_{\Sigma}d^{2}y\sqrt{\sigma}n^{a}%
\tau_{ab}\epsilon^{b}=\sigma\Omega^{1/2}f^{-1/2}\tau_{tt}\biggr{\vert}_{x_{b}%
}=\frac{4\pi}{\kappa}\biggl{[}\frac{\alpha+3\eta^{2}}{3\eta^{3}}%
+O(x-1)\biggr{]}
\end{equation}
where $n^{a}=(\partial_{t})^{a}/\sqrt{-g_{tt}}$ is the normal unit vector to
the surface $t=$constant.

We can also obtain the mass by using the method of \cite{Anabalon:2014fla}.
With the change of coordinates
\begin{equation}
x=1-\frac{1}{\eta r}+\frac{(\nu^{2}-1)}{24\eta^{3}r^{3}}\biggl{[}1+\frac
{1}{\eta r}-\frac{9(\nu^{2}-9)}{80\eta^{2}r^{2}}\biggr{]}+O(r^{-6})
\end{equation}
we can read off the mass from the subleading term of $g_{tt}$:
\begin{equation}
-g_{tt}=f(x)\Omega(x)=\frac{r^{2}}{l^{2}}+1+\frac{\alpha+3\eta^{2}}{3\eta
^{3}r}+O(r^{-3})\label{Lapse}%
\end{equation}
The reason is that the asymptotic expansion of the scalar field becomes in
these coordinates
\begin{equation}
\phi(x)=l_{\nu}^{-1}\ln{x}=-\frac{1}{l_{\nu}\eta r}-\frac{1}{2l_{\nu}\eta
^{2}r^{2}}+\frac{\nu^{2}-9}{24\eta^{3}r^{3}}+O(r^{-4})
\end{equation}
and we obtain that the coeficient of the leading term is $-l_{\nu}^{-1}%
\eta^{-1}$ and the subleading term is $-(2l_{\nu}\eta^{2})^{-1}$. Both modes
are normalizable and the \textit{conformal symmetry of the boundary is
preserved}. Therefore, we obtain
\begin{equation}
M=\frac{1}{2G}\biggl{(}\frac{\alpha+3\eta^{2}}{3\eta^{3}}\biggr{)}
\end{equation}
and this result matches the mass computed above with the quasilocal formalism.
Using the following expressions for the temperature and entropy
\begin{equation}
T=\frac{f^{^{\prime}}(x)}{4\pi\eta}\biggr{\vert}_{x=x_{h}}=\frac{1}{4\pi
\eta\Omega(x_{h})}\biggl{[}\frac{\alpha}{\eta^{2}}+2+\nu\frac{x_{h}^{\nu}%
+1}{x_{h}^{\nu}-1}\biggr{]}\,\,,\qquad S=\frac{A}{4G}=\frac{4\pi\Omega(x_{h}%
)}{4G}\label{temperature}%
\end{equation}
one can easily check that the first law $dM=TdS$ is satisfied.

As we have already mentioned, the solutions fall into two distinct classes.
For the the family with a positive scalar field the mass is
\begin{equation}
M=-\frac{1}{2G}\biggl{(}\frac{\alpha+3\eta^{2}}{3\eta^{3}}\biggr{)}
\end{equation}
and temperature
\begin{equation}
T=-\frac{1}{4\pi\eta\Omega(x_{h})}\biggl{[}\frac{\alpha}{\eta^{2}}+2+\nu
\frac{x_{h}^{\nu}+1}{x_{h}^{\nu}-1}\biggr{]}
\end{equation}
with the entropy given by the area law.

\section{Phase transitions}

The AdS spacetime can be thought to have a potential wall as one approaches
the asymptotic infinity and it behaves as an infinite box (more rigorously, it
has a conformal boundary). Since it is not a globally hyperbolic spacetime the
information can leak out or get in through the boundary and so, to obtain a
well defined problem, one has to impose boundary conditions.

The scalar field satisfies different boundary conditions depending on whether
it is positive or negative, which corresponds to the two families of solutions
mentioned before. A classical field theory is completely defined when the
boundary conditions are prescribed. For any boundary conditions on the scalar
field, the non-trivial vacuum configuration given by SAdS black hole solution
should be included as an allowed state of the theory. Hence, in the canonical
ensemble, its free energy can be compared with the one of the hairy black hole
at a given temperature. Figure $1(a)$ shows that, for the family with a
positive scalar field, SAdS is always more favorable than the hairy
configuration. Figure $1(b)$ shows the same phenomena for the family with a
negative scalar field. We found that generic values of $\alpha$ do not change
the qualitative behavior of the phase diagrams.
\begin{figure}[ptbh]
\hfill\begin{minipage}[h]{.45\textwidth}
\begin{center}
\epsfig{file=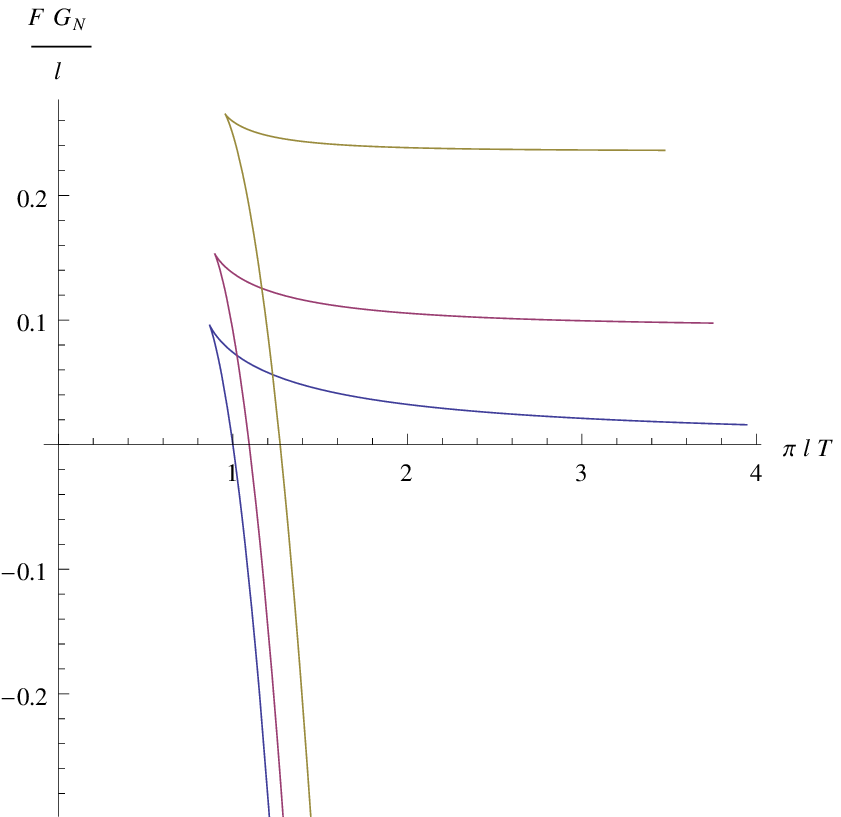,width=12cm,angle=0,scale=0.6}
\subfloat{(a)}
\label{fig-tc1}
\end{center}
\end{minipage}
\hfill\begin{minipage}[h]{.45\textwidth}
\begin{center}
\epsfig{file=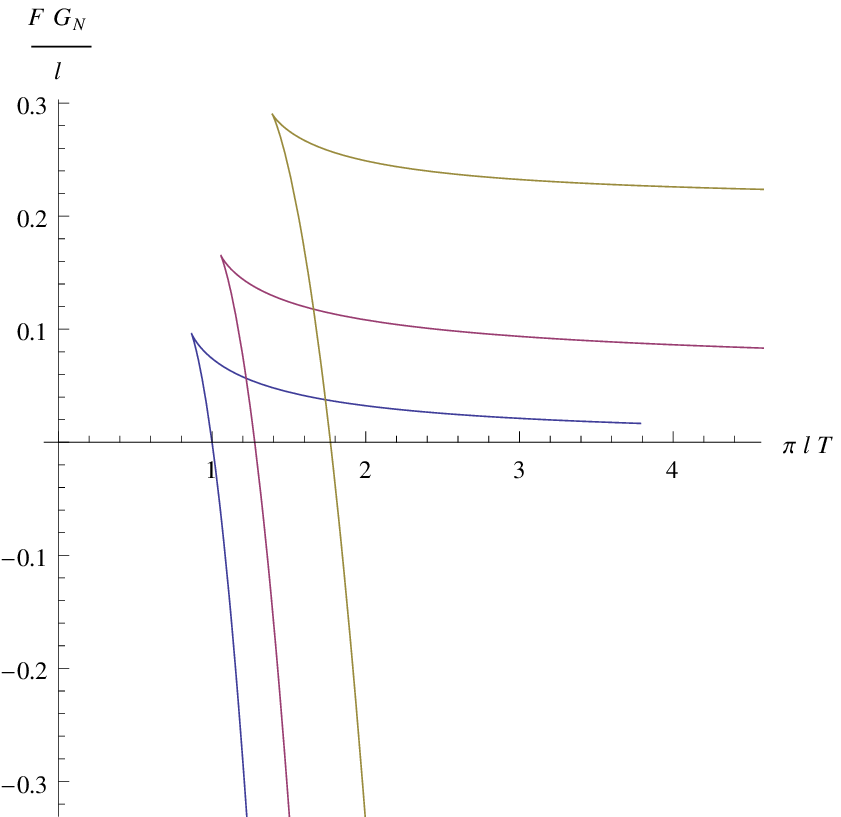,width=12cm,angle=0,scale=0.6}
\subfloat{(b)}
\label{fig-tc2}
\end{center}
\end{minipage}
\hfill\caption{(a) Dimensionless free energy versus dimensionless temperature,
for different values of $\nu$ and fixed $\alpha=-10 l^{-2}$ and positive
scalar field. The plots are for $\nu=1,\nu=1.9$ and $\nu=3$ (from down up).
The free energy of Schwarzschild AdS ($\nu=1$) goes to zero when T goes to
infinity. The free energy of the hairy black holes goes to a constant at
infinite temperature. (b) Dimensionless free energy versus dimensionless
temperature, for different values of $\nu$ and fixed $\alpha=10 l^{-2}$ and
negative scalar field. The plots are for $\nu=1,\nu=1.9$ and $\nu=3$ (from
down up). }%
\end{figure}

As in the SAdS case, there are two branches consisting of large and smaller
black holes. Figures $2(a)$ and $2(b)$ show the mass versus the temperature
for the families with a positive scalar field and a negative one,
respectively.
\begin{figure}[ptbh]
\hfill\begin{minipage}[h]{.45\textwidth}
\begin{center}
\epsfig{file=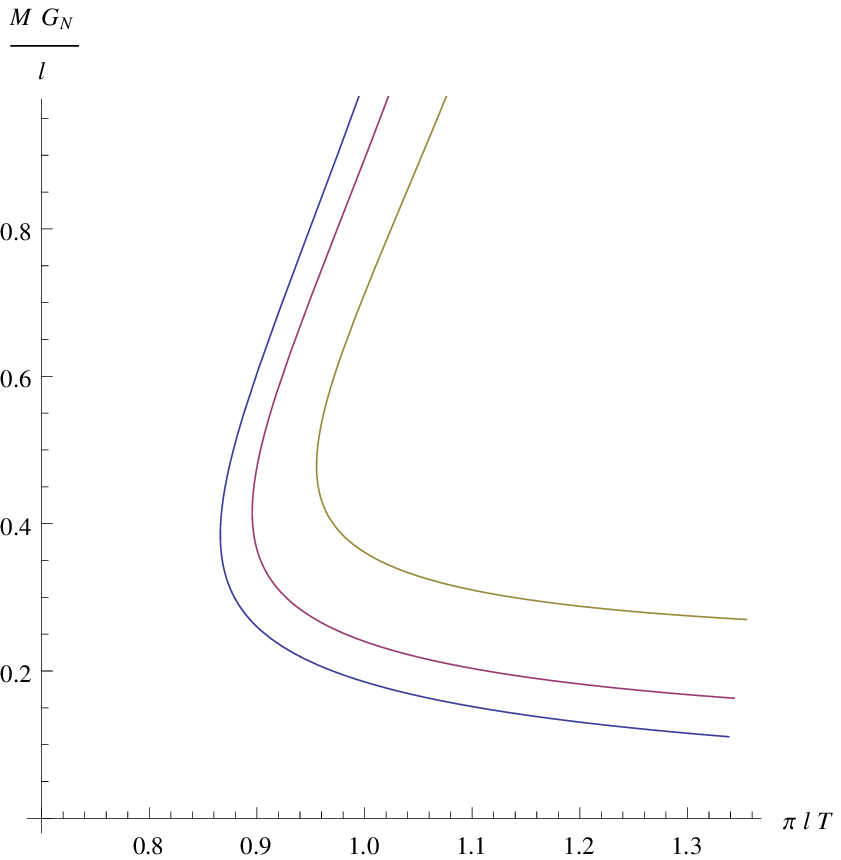,width=11cm,angle=0,scale=0.6}
\subfloat{(c)}
\label{fig-tc1}
\end{center}
\end{minipage}
\hfill\begin{minipage}[h]{.45\textwidth}
\begin{center}
\epsfig{file=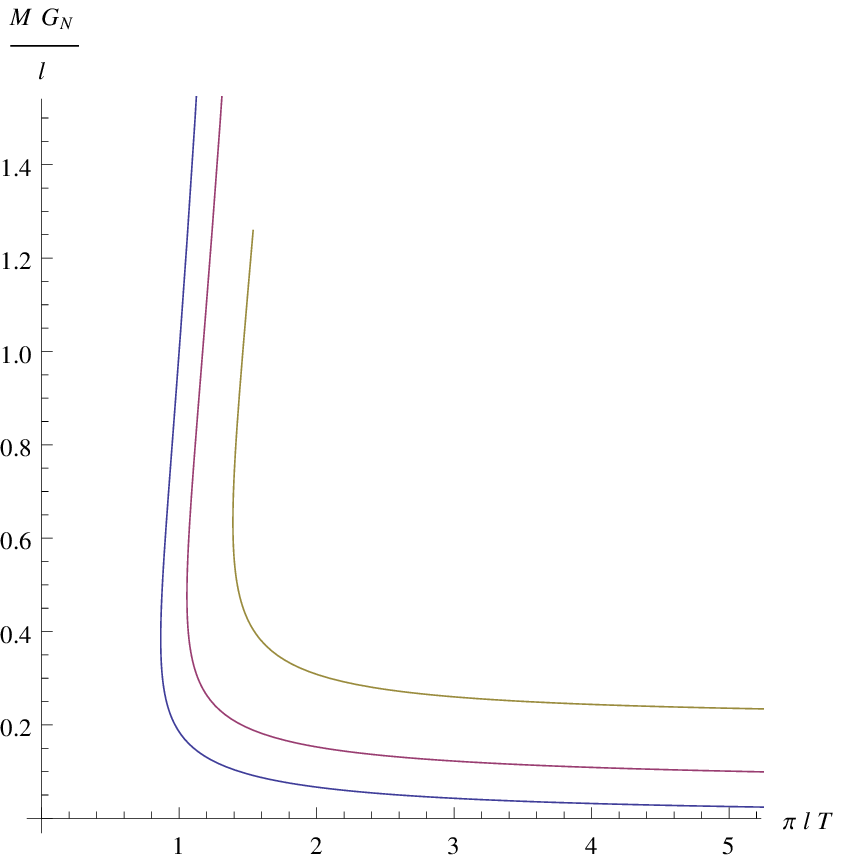,width=11cm,angle=0,scale=0.6}
\subfloat{(d)}
\label{fig-tc2}
\end{center}
\end{minipage}
\hfill\caption{(c) Dimensionless mass versus dimensionless temperature, for
different values of $\nu$ and fixed $\alpha=10 l^{-2}$ and negative scalar
field. The plots are for $\nu=1,\nu=1.9$ and $\nu=3$ (from left right). Here
it is possible to see the existence of small and large black hairy black holes
in exact resemblance with Schwarzschild AdS.\newline(d) Dimensionless mass
versus dimensionless temperature, for different values of $\nu$ and fixed
$\alpha=-10 l^{-2}$ and positive scalar field. The plots are for $\nu
=1,\nu=1.9$ and $\nu=3$ (from left right). Here it is possible to see the
existence of small and large black hairy black holes in exact resemblance with
Schwarzschild AdS.}%
\end{figure}
These plots provide information about the specific heat
\begin{equation}
C=\frac{\partial M}{\partial T}%
\end{equation}
that is interpreted as the slope.

The entire branch of smaller black holes (for both families) is unstable
thermodynamically and has a positive free energy, while the large black holes
branch are stable thermodynamically and the free energy goes negative for all
$T>T_{c}$. Unlike the planar black holes for which do not exist first order
phase transitions with respect to AdS, the free energy is changing the sign.
This is an indication that, for hairy black hole solutions with spherical
horizon geometry, there are first order phase transitions with respect to
thermal AdS --- the part of the branch of the large black holes that have a
negative free energy with respect to AdS are clearly the preferred ones.

\section{Conclusions}

From the point of view of AdS/CFT duality, the study of the thermodynamics of
asymptotically AdS black holes is relevant to understanding the phase diagram
of some holographic dual field theories. We have investigated the
thermodynamics of a general class of hairy black holes with boundary
conditions for a scalar field with the conformal mass $m=-2/l^{2}$, which
preserve the AdS isometries. It is worth remarking the close similarity that
we have observed with the familiar structure of SAdS black hole. The large
hairy black holes are thermodynamically stable, and the smaller ones have a
negative specific heat.

We have computed the Euclidean action (and so the thermodynamic potential) by
using the quasilocal formalism supplemented with counterterms. Using these
results, we have shown that there exist first order phase transitions between
the thermal AdS and hairy black hole. On the other hand, by comparing the free
energy of the hairy black hole with the one for the SAdS solution, it seems
that the SAdS black hole is always preferred.

An interesting future direction is to check the existence of gravitational
solitons as in \cite{Hertog:2004ns} and the implications for the phase
diagram. It will be also interesting to study the phase diagram of the family
of exact charged hairy black holes presented en \cite{Anabalon:2013sra}. In
this case, one can study both, the canonical and grand-canonical ensemble,
respectively. In the canonical ensemble the charge, which is an extensive
variable, should be kept fixed. Since AdS spacetime with a fixed charge is not
a solution of the equations of motion, it is appropriate to compute the
Euclidean action with respect to the ground state that is the extremal black
hole in this case \cite{Chamblin:1999hg}. Using similar arguments as in
\cite{Astefanesei:2011pz, Astefanesei:2010dk} it was shown in
\cite{Anabalon:2013sra} that there exist extremal black holes with a finite
horizon area and so it is expected that the canonical ensemble is well defined.

\vskip 1cm

\section*{Acknowledgments}

We would like to thank Cristian Martinez for interesting discussions and
collaboration on related projects and Wellington Galleas for useful
conversations. Research of A.A. is supported in part by the Fondecyt Grants N%
${{}^o}$
11121187, 1141073 and by the Newton-Picarte Grants DPI20140053 and
DPI20140115. Research of D.A. has been partially funded by the Fondecyt grants
1120446 and by the Newton-Picarte Grant DPI20140115. Research of D.C. is
supported in part by the CONICYT Ph.D. scholarship and the Newton-Picarte
Grant DPI20140115 and DPI20140053.

\end{document}